\input harvmac

\let\includefigures=\iftrue
\let\useblackboard=\iftrue
\newfam\black

\includefigures
\message{If you do not have epsf.tex (to include figures),}
\message{change the option at the top of the tex file.}
\input epsf
\def\figin{\epsfcheck\figin}\def\figins{\epsfcheck\figins}
\def\epsfcheck{\ifx\epsfbox\UnDeFiNeD
\message{(NO epsf.tex, FIGURES WILL BE IGNORED)}
\gdef\figin##1{\vskip2in}\gdef\figins##1{\hskip.5in}
\else\message{(FIGURES WILL BE INCLUDED)}%
\gdef\figin##1{##1}\gdef\figins##1{##1}\fi}
\def\DefWarn#1{}
\def\figinsert{\goodbreak\midinsert}
\def\ifig#1#2#3{\DefWarn#1\xdef#1{fig.~\the\figno}
\writedef{#1\leftbracket fig.\noexpand~\the\figno}%
\figinsert\figin{\centerline{#3}}\medskip\centerline{\vbox{
\baselineskip12pt\advance\hsize by -1truein
\noindent\footnotefont{\bf Fig.~\the\figno:} #2}}
\bigskip\endinsert\global\advance\figno by1}
\else
\def\ifig#1#2#3{\xdef#1{fig.~\the\figno}
\writedef{#1\leftbracket fig.\noexpand~\the\figno}%
\global\advance\figno by1}
\fi
%

\useblackboard
\message{If you do not have msbm (blackboard bold) fonts,}
\message{change the option at the top of the tex file.}
\font\blackboard=msbm10 scaled \magstep1
\font\blackboards=msbm7
\font\blackboardss=msbm5
\textfont\black=\blackboard
\scriptfont\black=\blackboards
\scriptscriptfont\black=\blackboardss

\else

\fi
%
\def\subsubsec#1{\bigskip\noindent{\it{#1}} \bigskip}
\def\yboxit#1#2{\vbox{\hrule height #1 \hbox{\vrule width #1
\vbox{#2}\vrule width #1 }\hrule height #1 }}
\def\fillbox#1{\hbox to #1{\vbox to #1{\vfil}\hfil}}
\def\ybox{{\lower 1.3pt \yboxit{0.4pt}{\fillbox{8pt}}\hskip-0.2pt}}
%
%


\def\comments#1{}

\def\p{\partial}

\def\half{{1\over 2}}

\def\tr{{\rm tr }}

\def\vev#1{\langle{#1}\rangle}

\def\CK{{\cal K}}

\def\CN{{\cal N}}
\def\CO{{\cal O}}

\def\CL{{\cal L}}

\def\CX{{\cal X}}

\def\a{\alpha}

\def\II{\relax{I\kern-.10em I}}

%

\def\IZ{\relax{\rm Z\kern-.34em Z}}
\def\IB{\relax{\rm I\kern-.18em B}}
\def\IC{{\relax\hbox{$\inbar\kern-.3em{\rm C}$}}}
\def\ID{\relax{\rm I\kern-.18em D}}
\def\IE{\relax{\rm I\kern-.18em E}}
\def\IF{\relax{\rm I\kern-.18em F}}
\def\IG{\relax\hbox{$\inbar\kern-.3em{\rm G}$}}
\def\IGa{\relax\hbox{${\rm I}\kern-.18em\Gamma$}}
\def\IH{\relax{\rm I\kern-.18em H}}
\def\II{\relax{\rm I\kern-.18em I}}
\def\IK{\relax{\rm I\kern-.18em K}}
\def\IP{\relax{\rm I\kern-.18em P}}

%

\def\inbar{\,\vrule height1.5ex width.4pt depth0pt}

\def\p{\partial}

\def\wdg{\wedge}

\def\k{\kappa}

\def\IR{\relax{\rm I\kern-.18em R}}


\def\simlt{\hskip0.05in\relax{
\raise3.0pt\hbox{ $<$
{\lower5.0pt\hbox{\kern-1.05em $\sim$}} }} \hskip0.05in}
\def\simgt{\hskip0.05in\relax{
\raise3.0pt\hbox{ $>$
{\lower5.0pt\hbox{\kern-1.05em $\sim$}}}} \hskip0.05in}

\def\Jb {{\bar{J}}}

%


%

\def\lp10{\ell_p^{10}}
\def\lp11{\ell_p^{11}}
\def\R11{R_{11}}

\def\frac#1#2{{#1 \over #2}}

\def\ch{{\rm ch}}
\def\dg{{\dagger}}
\def\td{{\rm Td}}



\newdimen\tableauside\tableauside=1.0ex
\newdimen\tableaurule\tableaurule=0.4pt
\newdimen\tableaustep
\def\phantomhrule#1{\hbox{\vbox to0pt{\hrule height\tableaurule width#1\vss}}}
\def\phantomvrule#1{\vbox{\hbox to0pt{\vrule width\tableaurule height#1\hss}}}
\def\sqr{\vbox{%
  \phantomhrule\tableaustep
  \hbox{\phantomvrule\tableaustep\kern\tableaustep\phantomvrule\tableaustep}%
  \hbox{\vbox{\phantomhrule\tableauside}\kern-\tableaurule}}}
\def\squares#1{\hbox{\count0=#1\noindent\loop\sqr
  \advance\count0 by-1 \ifnum\count0>0\repeat}}
\def\tableau#1{\vcenter{\offinterlineskip
  \tableaustep=\tableauside\advance\tableaustep by-\tableaurule
  \kern\normallineskip\hbox
    {\kern\normallineskip\vbox
      {\gettableau#1 0 }%
     \kern\normallineskip\kern\tableaurule}%
  \kern\normallineskip\kern\tableaurule}}
\def\gettableau#1 {\ifnum#1=0\let\next=\null\else
  \squares{#1}\let\next=\gettableau\fi\next}

\tableauside=1.0ex
\tableaurule=0.4pt


 %
 %
 \def\eqnn#1{\xdef #1{(\secsym\the\meqno)}\writedef{#1\leftbracket#1}%
 \global\advance\meqno by1\wrlabeL#1}
 \def\eqna#1{\xdef #1##1{\hbox{$(\secsym\the\meqno##1)$}}
 \writedef{#1\numbersign1\leftbracket#1{\numbersign1}}%
 \global\advance\meqno by1\wrlabeL{#1$\{\}$}}
 \def\eqn#1#2{\xdef #1{(\secsym\the\meqno)}\writedef{#1\leftbracket#1}%
 \global\advance\meqno by1$$#2\eqno#1\eqlabeL#1$$}

\global\newcount\itemno \global\itemno=0

\def\itemaut#1{\global\advance\itemno by1\noindent\item{\the\itemno.}#1}


\def\eg{{\it e.g.}}
\def\ie{{\it i.e.}}

\hyphenation{Di-men-sion-al}



\lref\DineXK{
  M.~Dine, N.~Seiberg and E.~Witten,
 ``Fayet-Iliopoulos Terms in String Theory,''
  Nucl.\ Phys.\ B {\bf 289}, 589 (1987).
}

\lref\IbanezQP{
  L.~E.~Ibanez, R.~Rabadan and A.~M.~Uranga,
 ``Anomalous $U(1)$'s in Type I and Type IIB $D = 4$, ${\cal N} = 1$ string vacua,''
  Nucl.\ Phys.\ B {\bf 542}, 112 (1999), hep-th/9808139.
}

\lref\IntriligatorAW{
  K.~Intriligator and N.~Seiberg,
 ``The Runaway Quiver,'' hep-th/0512347.
}

\lref\PoppitzWP{
  E.~Poppitz, Y.~Shadmi and S.~P.~Trivedi,
 ``Supersymmetry Breaking and Duality in $SU(N)\times SU(N-M)$ Theories,''
  Phys.\ Lett.\ B {\bf 388}, 561 (1996),hep-th/9606184.
}

\lref\WittenGLSM{E. Witten, ``Phases of ${\cal N}=2$ Theories in Two Dimensions,''
Nucl. Phys. B {\bf 403} (1993) 159, hep-th/9301042.}

\lref\stringinst{
E. Witten, ``Small Instantons in String Theory,'' Nucl. Phys. B {\bf 460}
(1996) 541, :hep-th/9511030;
M. Douglas, ``Branes within Branes,'' hep-th/9512077.}

\lref\PoppitzVH{
  E.~Poppitz, Y.~Shadmi and S.~P.~Trivedi,
  ``Duality and Exact Results in Product Group Theories,''
  Nucl.\ Phys.\ B {\bf 480}, 125 (1996), hep-th/9605113.
}

\lref\KST{S. Kachru, M. Schulz and S. Trivedi, ``Moduli Stabilization from
Fluxes in a Simple IIB Orientifold,'' JHEP {\bf 0310} (2003) 007, hep-th/0201028.}

\lref\Polch{J. Polchinski, {\it String Theory}, vol.2, Cambridge University Press.}

\lref\ADS{
I. Affleck, M. Dine and N. Seiberg, ``Supersymmetry Breaking in
Supersymmetric QCD,"  Nucl. Phys. B {\bf 241} (1984) 493.}

\lref\AffleckXZ{
  I.~Affleck, M.~Dine and N.~Seiberg,
 ``Dynamical Supersymmetry Breaking in Four Dimensions and its
 Phenomenological Implications,''
  Nucl.\ Phys.\ B {\bf 256}, 557 (1985).
}

\lref\Bertolini{
R. Argurio, M. Bertolini, C. Closset and S. Cremonesi, ``On Stable
Nonsupersymmetric Vacua at the Bottom of Cascading Theories,'' hep-th/0606175.}

\lref\ArkaniHamedNU{
  N.~Arkani-Hamed, M.~Dine and S.~P.~Martin,
 ``Dynamical Supersymmetry Breaking in Models with a Green-Schwarz
 Mechanism,''
  Phys.\ Lett.\ B {\bf 431}, 329 (1998), hep-ph/9803432.
}

\lref\BerkoozKM{
  M.~Berkooz, M.~R.~Douglas and R.~G.~Leigh,
 ``Branes Intersecting at Angles,''
  Nucl.\ Phys.\ B {\bf 480}, 265 (1996), hep-th/9606139.
}
\lref\VP{L.~ Martucci, J.~ Rosseel, D.~ Van de Bleeken, A.~ Van Proeyen,
``Dirac Actions for D-Branes on Backgrounds with Fluxes'', hep-th/0504041.}
\lref\Saul{N.~Saulina, ``Topological Constraints on Stabilized Flux Vacua'',
Nucl. Phys. B {\bf 720} (2005) 203-210, hep-th/0503125.}
\lref\Kalosh{R.~Kallosh, A.-K. Kashani-Poor, A.~ Tomasiello,
``Counting fermionic zero modes on M5 with fluxes'',hep-th/0503138,
 JHEP {\bf 0506} (2005) 069.}

\lref\Kaloshii{ E.~Bergshoeff, R.~ Kallosh, A.-K. Kashani-Poor, D.~ Sorokin, A.~ Tomasiello,
``An Index for the Dirac Operator on D3-Branes with Background Fluxes'',
hep-th/0507069, JHEP {\bf 0510} (2005) 102.}

\lref\BerensteinXA{
  D.~Berenstein, C.~P.~Herzog, P.~Ouyang and S.~Pinansky,
 ``Supersymmetry Breaking from a Calabi-Yau Singularity,''
  JHEP {\bf 0509}, 084 (2005), hep-th/0505029.
}

\lref\BertoliniDI{
  M.~Bertolini, F.~Bigazzi and A.~L.~Cotrone,
 ``Supersymmetry Breaking at the End of a Cascade of Seiberg Dualities,''
  Phys.\ Rev.\ D {\bf 72}, 061902 (2005), hep-th/0505055.
}

\lref\FrancoZU{
  S.~Franco, A.~Hanany, F.~Saad and A.~M.~Uranga,
 ``Fractional Branes and Dynamical Supersymmetry Breaking,''
  JHEP {\bf 0601}, 011 (2006), hep-th/0505040.
}

\lref\WittenNF{
  E.~Witten,
 ``Dynamical Breaking of Supersymmetry,''
  Nucl.\ Phys.\ B {\bf 188}, 513 (1981).
}

\lref\WittenXY{
  E.~Witten,
 ``Baryons and Branes in anti de Sitter Space,''
  JHEP {\bf 9807}, 006 (1998), hep-th/9805112.
}

\lref\Bershadsky{M. Bershadsky et al, ``F-theory, Geometric
Engineering and ${\cal N}=1$ Dualities,'' Nucl. Phys. B {\bf 505} (1997) 153, hep-th/9612052.}

\lref\GanorPE{
  O.~J.~Ganor,
 ``A Note on Zeroes of Superpotentials in F-theory,''
  Nucl.\ Phys.\ B {\bf 499}, 55 (1997), hep-th/9612077.
}

\lref\BeasleyXV{
  C.~E.~Beasley,
 ``BPS Branes from Baryons,''
  JHEP {\bf 0211}, 015 (2002), hep-th/0207125.
}

\lref\MikhailovYA{
  A.~Mikhailov,
 ``Giant Gravitons from Holomorphic Surfaces,''
  JHEP {\bf 0011}, 027 (2000), hep-th/0010206.
}

\lref\DiaconescuPC{
  D.-E.~Diaconescu, B.~Florea, S.~Kachru and P.~Svrcek,
 ``Gauge-Mediated Supersymmetry Breaking in String Compactifications,''
  JHEP {\bf 0602}, 020 (2006), hep-th/0512170.
}

\lref\WittenEG{
  E.~Witten,
 ``World-sheet Corrections via D-Instantons,''
  JHEP {\bf 0002}, 030 (2000), hep-th/9907041.
}

\lref\WittenBN{
  E.~Witten,
 ``Non-Perturbative Superpotentials in String Theory,''
  Nucl.\ Phys.\ B {\bf 474}, 343 (1996), hep-th/9604030.
}

\lref\PolchinskiMT{
  J.~Polchinski,
  ``Dirichlet-Branes and Ramond-Ramond Charges,''
  Phys.\ Rev.\ Lett.\  {\bf 75}, 4724 (1995), hep-th/9510017.
}

\lref\VafaQF{
  C.~Vafa,
  ``Brane/Antibrane Systems and $U(N|M)$ Supergroup,'' hep-th/0101218.
}

\lref\KachruAW{
  S.~Kachru, R.~Kallosh, A.~Linde and S.~P.~Trivedi,
``de Sitter Vacua in String Theory,''
  Phys.\ Rev.\ D {\bf 68}, 046005 (2003), hep-th/0301240.
}

\lref\Examples{
F. Denef, M. Douglas and B. Florea, JHEP {\bf 0406} (2004) 034, hep-th/0404257;
F. Denef, M. Douglas, B. Florea, A. Grassi and S. Kachru, hep-th/0503124;
D. Lust, S. Reffert, W. Schulgin and S. Stieberger, hep-th/0506090;
D.~Lust, S.~Reffert, E.~Scheidegger, W.~Schulgin and S.~Stieberger, hep-th/0609013.
}
\lref\BalasubramanianZX{
  V.~Balasubramanian, P.~Berglund, J.~P.~Conlon and F.~Quevedo,
   ``Systematics of Moduli Stabilisation in Calabi-Yau Flux
 Compactifications,''
  JHEP {\bf 0503}, 007 (2005), hep-th/0502058.
}

\lref\JockersZY{
  H.~Jockers and J.~Louis,
 ``D-Terms and F-Terms from D7-Brane Fluxes,''
  Nucl.\ Phys.\ B {\bf 718}, 203 (2005), hep-th/0502059.
}

\lref\Ouyang{I. Klebanov, P. Ouyang and E. Witten, ``A Gravity Dual of
the Chiral Anomaly,'' Phys. Rev. D {\bf 65} (2002) 105007, hep-th/0202056.}

\lref\Ejaz{C. Herzog, Q. Ejaz and I. Klebanov, ``Cascading RG Flows
from New Sasaki-Einstein Manifolds,'' JHEP {\bf 0502} (2005) 009, hep-th/0412193.}

\lref\NekrasovJS{
  N.~Nekrasov, H.~Ooguri and C.~Vafa,
   ``S-Duality and Topological Strings,''
   JHEP {\bf 0410}, 009 (2004), hep-th/0403167.}

\lref\fulton{ W.~Fulton, {\it Introduction to Toric Varieties }, Princeton University Press.}

\lref\stringstalk{
Talk given by J. McGreevy at Strings 2006, Beijing,
``On the Capture of Runaway Quivers,"
{\tt http://strings06.itp.ac.cn/talk-files/mcgreevy.pdf}.
}

\lref\cvetic{R. Blumenhagen, M. Cvetic, T. Weigand, hep-th/0609191.}
\lref\lust{M. Haack, D. Krefl, D. L\"ust, A. Van Proeyen, and M. Zagermann,
hep-th/0609211.}

\lref\urangayesterday{R. Ibanez and A. Uranga, hep-th/0609213.}
\lref\hermanetal{M. Buican, D. Malyshev, D. R. Morrison, H. Verlinde, M. Wijnholt,
``D-Branes at Singularities, Compactification, and Hypercharge,"
hep-th/0610007.}

\lref\VerlindeJR{
  H.~Verlinde and M.~Wijnholt,
 ``Building the Standard Model on a D3-Brane,''
  arXiv:hep-th/0508089.
}
\lref\Preskill{
  J.~Preskill,
 ``Gauge Anomalies in an Effective Field Theory,''
  Annals Phys.\  {\bf 210}, 323 (1991).
}

\lref\BerensteinFI{
  D.~Berenstein and M.~R.~Douglas,
 ``Seiberg Duality for Quiver Gauge Theories,'' hep-th/0207027.
}

\lref\Moorei{
M. R. Douglas and G. W. Moore, ``D-Branes, Quivers, and ALE Instantons,'' hep-th/9603167.}

\lref\Mooreii{
R. Minasian and G. W. Moore, ``K-theory and Ramond-Ramond Charge,''
JHEP {\bf 9711}, 002 (1997), hep-th/9710230.}

\Title{\vbox{\baselineskip12pt\hbox{hep-th/0610003}
\hbox{MIT-CTP/3774}
\hbox{SU-ITP-06/26}
\hbox {SLAC-PUB-12138}}}
{\vbox{
\centerline{Stringy Instantons and Quiver Gauge Theories}}}
\bigskip
\centerline{Bogdan Florea$^1$, Shamit Kachru$^{1,2}$, John McGreevy$^{1,3}$, and
Natalia Saulina$^{4,5}$}
\bigskip
\centerline{$^1${\it Department of Physics and SLAC,
Stanford University, Palo Alto, CA 94305-4060}}
\bigskip
\centerline{$^2${\it Kavli Institute for Theoretical Physics,
University of California, Santa Barbara, CA 93106-4030}}
\bigskip
\centerline{$^3${\it Center for Theoretical Physics,
Massachusetts Institute of Technology, Cambridge, MA 02139}}
\bigskip
\centerline{$^4${\it Department of Physics,
Harvard University, Cambridge, MA 02138}}
\bigskip
\centerline{$^5${\it Department of Physics,
California Institute of Technology, Pasadena, CA 91125}}
\bigskip
\bigskip
\noindent
We explore contributions to the $4D$ effective superpotential
which arise from Euclidean
D3 branes (``instantons'')
that intersect space-filling D-branes.
These effects can perturb the effective field theory on the space-filling
branes by nontrivial operators composed of charged matter fields,
changing the vacuum structure in a qualitative
way in some examples.
Our considerations are exemplified throughout by a careful study of a
fractional brane configuration on a del Pezzo surface.

\bigskip
\Date{September 2006}

\newsec{Introduction}

Models of dynamical supersymmetry breaking are of great interest, both
for the intricate dynamics they exhibit, and for their possible application
to the hierarchy problem \WittenNF.
By definition, such models invoke non-perturbative effects in 4d
supersymmetric quantum field theory.  In the context of string theory,
where the 4d low-energy supersymmetric theory arises via compactification,
one expects that instanton effects in the field theories on space-filling
D-branes
can be recast as (Euclidean) D-brane instanton
effects \stringinst.
Perhaps the simplest example is the theory on $N$ D3 branes; the Yang-Mills
instanton in this theory is simply the D$(-{\rm 1})$ brane inside the D3 branes.

More generally, other Euclidean D-branes which ${\it cannot}$
be interpreted as instantons in non-Abelian gauge groups can
also contribute to
the superpotential, and hence may be important in the study of supersymmetry
breaking.  For instance in F-theory on an elliptic
Calabi-Yau fourfold $X \to B$,
Witten argued some time ago \WittenBN\
that contributions to the superpotential for
K\"ahler moduli arise from
D3's which wrap divisors in $B$ that lift to vertical divisors
of holomorphic Euler character $\chi = 1$ in $X$.
These effects, though non-perturbative in strength, can play a
crucial role in low-energy physics.  For example, they
are useful in constructing controlled examples of moduli (meta)stabilization
\refs{\KachruAW,\Examples,\BalasubramanianZX}.

In cases where the instanton intersects other (space-filling) D-branes,
the analysis of \WittenBN\ does not immediately apply.
It is then useful to
find
techniques to
compute directly the superpotential in this situation.
For instance, suppose that one has arranged for a nontrivial
quiver field theory to arise at
a singularity in $B$.
That is, one has some nontrivial spectrum of D3, D5 and D7 branes wrapping
various cycles in the singular geometry.  Then Euclidean D3's which
intersect the singularity, will in general
have massless strings which stretch to
the occupied nodes of the quiver (as was first discussed in some simple
cases by Ganor \GanorPE).
These modes will couple to the quiver fields, and the integral over
these collective coordinates will then pull down nontrivial operators
involving fields in the quiver gauge theory.  Schematically, if the Euclidean
D3 wraps a cycle of volume $V$, and the quiver theory has charged fields
$\phi_i$, one finds effects of the form
\eqn\schemop{
\Delta W \sim {\cal O}(\phi_i) {\rm Exp}(-V)~.}
We will see that the operators ${\cal O}$ generated in this way
can have a qualitative effect on the spacetime physics.

The K\"ahler moduli chiral multiplets actually play a
second, equally important role in these theories.  The quiver gauge
theories generically come with some number of anomalous $U(1)$ factors.
The anomalies are cancelled by the Green-Schwarz mechanism, whereby
certain RR axions shift under the anomalous $U(1)$ \refs{\DineXK,\Moorei,\IbanezQP}.
These axions fill out (complex) scalar components of chiral multiplets
in the supersymmetric effective theory; they are the partners of the
K\"ahler moduli.
Therefore, for K\"ahler moduli partnered with the charged axions, the
associated instanton
action is no longer gauge invariant!  However,
the relevant contributions \schemop\ are generated in just such a way
that the operator ${\cal O}$ carries opposite charge under all
$U(1)$ symmetries, yielding a gauge invariant contribution to $W$.

In this paper, we initiate the systematic study of such instanton effects.
We focus
on a particular example, the fractional brane in the Calabi-Yau cone over
$dP_1$.
In addition to providing an ideal expository example, this theory is of
independent physical interest.
It was conjectured to dynamically break supersymmetry
in the pioneering works
\refs{\BerensteinXA,\FrancoZU,\BertoliniDI}.
The thorough analysis in \IntriligatorAW\ proved that in a suitable
decoupling limit,
the theory
has no stable vacuum.
These authors also stressed that various possible perturbations of
the superpotential could stabilize the model, as was further explored
for a wide class of models in the recent work \Bertolini.
The decoupling limit of \IntriligatorAW\ involves taking the limit
$M_X \to \infty$, where $M_X$ is the mass of
the anomalous $U(1)$ gauge bosons.
In what follows, we will see that this is a subtle
limit, since for any finite but large $M_X$ the behavior of the
theory
is qualitatively different.  In particular, finite $M_X$ will arise
in the compactified realizations of these models \DiaconescuPC, assuming
moduli can be stabilized.
We will briefly discuss moduli stabilization in such examples at the end of
this paper.
A similar analysis can be carried out
for many other quiver gauge theories;
we leave this for the future.

The organization of the paper is as follows.  In \S2, we describe our
main example, the quiver gauge theory arising from the Calabi-Yau cone
over $dP_1$.  Here we fix notation, describe the exceptional collection
and various mathematical facts which we will use in our calculations, and
discuss basic aspects of the fractional brane gauge theory.
The field theory we find in \S2\ has two anomalous $U(1)$ factors.
In \S3, we first briefly discuss why it is appropriate to keep these
factors (and their associated D-terms) in the analysis of the vacuum
structure of the compactified string theory.
We then review the relevant Green-Schwarz mechanism, and compute the charges
of the various axions under the anomalous $U(1)$'s.
In \S4, we
compute the numbers of strings stretching from a Euclidean D3
to a D7, in terms of dimensions of appropriate cohomology groups.
We then apply this
to give concrete results for the spectrum of strings
between a Euclidean D3 wrapping the $dP_1$
(with some general line bundle on its worldvolume), and the nodes of
the quiver.
In \S5, we spell out the rules for instantons to contribute,
and discuss two contributing instantons.  One can be
straightforwardly interpreted as the Affleck-Dine-Seiberg instanton
in one of the quiver nodes, and contributes the expected field theory
superpotential \ADS.  The other is an instanton which does not lie entirely
in any of the nodes of the quiver, and is in this sense a ``stringy''
effect which depends on the UV completion.  It gives rise to a baryonic
perturbation of the superpotential that qualitatively changes the
low-energy physics.  Finally, in \S6, we discuss the vacuum structure
of the resulting compactified quiver theory, including both the new
instanton effect and the effect of the $U(1)$ D-terms.

The effects we describe occur in any string compactification
containing space-filling branes.
For example, the contributions of D-instantons
are also modified when the Standard Model lives on such D-branes,
as in \eg\ \VerlindeJR.  The possibility that $\mu$-terms are generated
in this way has been pursued by \hermanetal.
Very recently \urangayesterday, such effects were applied
to neutrino masses in a similar context.
While we were brooding over the
selection rules implied by the R-anomaly,
we also learned of the work \refs{\cvetic, \lust},
which has some overlap with ours.

\newsec{The $dP_1$ Quiver}

We will study branes in the complex cone over $dP_1$ (or equivalently
$\IF_1$).  Since $dP_1$ has
$h^{0,0}=h^{2,2}=1$ and $h^{1,1}=2$, a general brane is specified by four
charges and the quiver will have four nodes.
Denote the nontrivial two-cycles in $\IF_1$ (the fiber and the base, viewing
$\IF_1$ as a $\IP^1$ bundle over $\IP^1$) as $f$ and $C_0$.  Then
a useful basis of branes is
given by:
\eqn\Coll{ \{ \CL_1, \CL_2, \CL_3, \CL_4 \} =
\{ \CO_{\IF_1},
\CO_{\IF_1}(C_0 + f),
\overline{\CO_{\IF_1}(f)},
\overline{ \CO_{\IF_1}(C_0)}
\} .}

This quiver has one non-anomalous fractional brane, given by the
combination
\eqn\fracguy{[D5] = \CL_1 + 2 \CL_2 + 3 \CL_3~.}
In addition, one can take any multiple $M \geq 1$ of this representation.
The quiver summarizing the field content for general $M$ is given below.
We will focus on the case $M=1$ for the most part.
\ifig\contentiousquiver{
The quiver for the nonanomalous fractional brane wrapping $dP_1$.}
{\epsfxsize2.0in\epsfbox{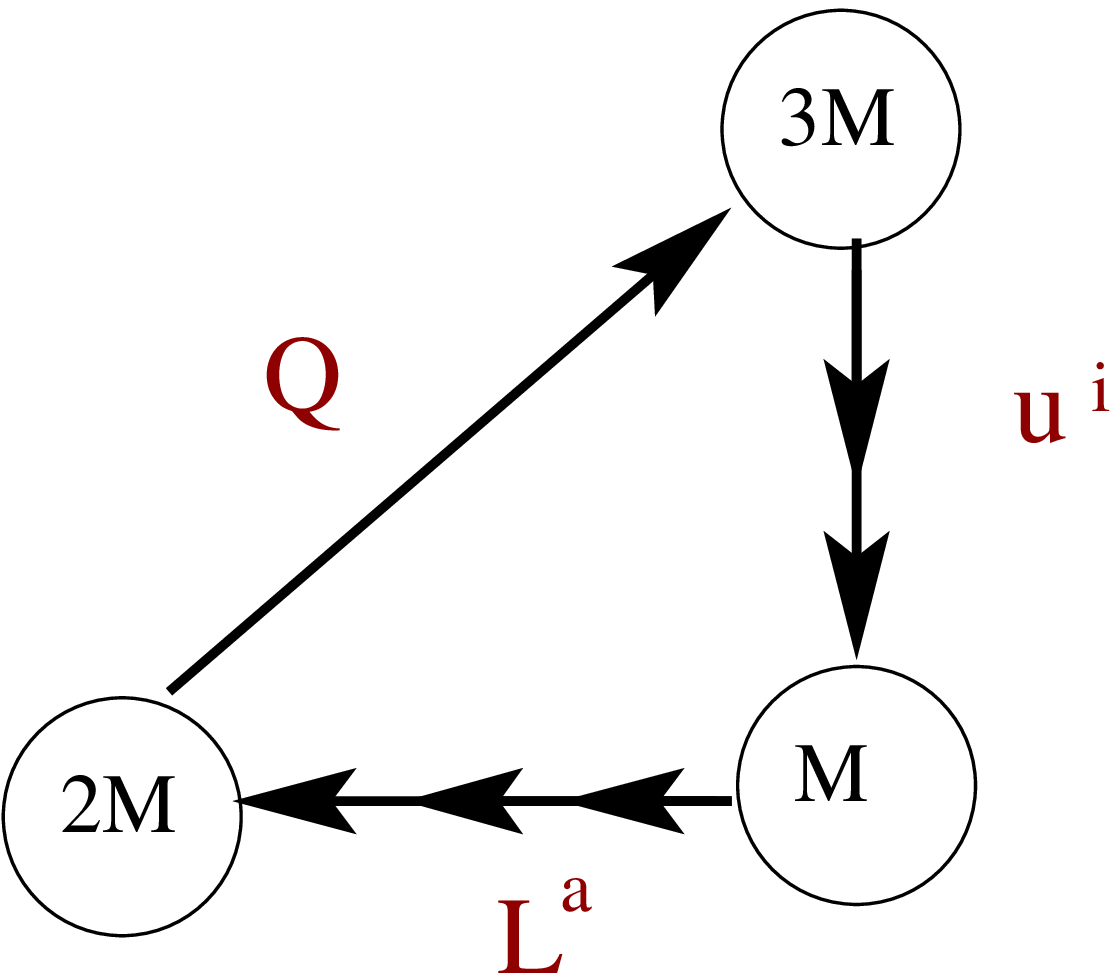}}

The theory on $M$ fractional branes is characterized by
gauge group $U(3M) \times U(2M) \times U(M)$.
We will denote the corresponding $U(1)$ factors as $U(1)_3$, $U(1)_2$ and
$U(1)_1$.  The sum of the three $U(1)$'s decouples (all fields are neutral),
while the two remaining combinations are in fact anomalous.

In an approximation where one discards the anomalous $U(1)$'s and their
D-terms, the dynamics of the field theory with field content as in the
quiver above and tree-level superpotential
\eqn\treelvl{W_{tree} = \lambda Q\epsilon_{ij}u^iL^j}
(which preserves a global $SU(2)_F$ flavor symmetry)
was analyzed in detail in
\IntriligatorAW.
For convenience, we reproduce here the table from \IntriligatorAW\ summarizing
the gauge and global symmetry representations of the various fields:
\eqn\chargesum{\matrix{&U(3M)&U(2M)&U(M)&[ SU(2)_F & U(1)_F & U(1)_R ]\cr \cr
Q&{\bf 3M}&{\bf{\overline{2M}}}&{\bf 1}&{\bf 1}&1&-1\cr\cr
 u&{\bf{\overline{3M}}}&{\bf 1}&{\bf M}&{\bf 2}&-1&0 \cr \cr
L&{\bf 1}&{\bf 2M}&{\bf{\overline{M}}}&{\bf 2}&0&3\cr\cr
L_3&{\bf 1}&{\bf 2M}&{\bf{\overline{M}}}&{\bf 1}&-3&-1,}}
where the groups in the brackets are global symmetries of the theory.

The $U(1)_R$  symmetry has triangle anomalies with
the $U(1)_B$ baryonic gauge symmetries
\eqn\ranomaly{{\cal A}\bigl(U(1)_R U(1)_a U(1)_b\bigr)=
-2r_a\,r_b. }
Here $r_a=(1,2,-3)$ is the rank of the a-th node of the quiver.
When the quiver gauge theory is realized in terms of D-branes
in the complex cone over $\IF_1,$ the R-anomaly can be cured\foot{In section 3 we consider compact model which involves an orientifold
of IIB string theory and projects RR 2-form in $M^4$ out of the massless fields.
Therefore, in the compact model the R-symmetry is broken.
It must also be broken perturbatively in the compact model
since such isometry would imply reduced holonomy.
Note that this is particularly important in the $M>1$ case
where \Bertolini\ has shown that no gauge invariant
operator has the right R-symmetry to appear in the superpotential.}
by the coupling of the gauge fields to the RR 2-form in $\IR^4$ as shown in
the Appendix.

Consider the $M=1$ case in the limit $\Lambda_3 \gg \Lambda_2$, so that the $SU(2)$ gauge
factor can
be treated as a global symmetry in the leading approximation.
The $SU(3)$ factor effectively has $N_f = 2 = N_c-1$, and so its dynamics
generate a non-perturbative correction to \treelvl\ -- the Affleck-Dine-Seiberg
superpotential \ADS:
\eqn\nonpW{W_{quantum} = W_{tree} + {\Lambda_3^7 \over {\rm det}(Q\cdot u)}~.}
The theory has no supersymmetric vacua, in keeping
with the arguments of
\refs{\BerensteinXA,\FrancoZU,\BertoliniDI}.  However, it has no
stable vacuum at all \IntriligatorAW.

We will see that the compactified string theory modifies the story in two
interesting ways.  Firstly,
when the string scale is finite, the anomalous $U(1)_B$s
do not decouple, and their anomalies
are cancelled via the Green-Schwarz mechanism.
Secondly, the
quantum superpotential \nonpW\ can receive further contributions from
``stringy'' effects, that modify the story in an interesting way.

\newsec{Anomalous $U(1)$'s}

\subsec{Generalities}

In this subsection we briefly review the effective field theory
of anomalous $U(1)$'s
\refs{\DineXK, \Preskill, \BerkoozKM, \IbanezQP}.  In the next, we
compute the axion charges
required to cancel anomalies in our theory.
In many ways, the relevant Green-Schwarz mechanism is a fancy string theory
realization of
a simple field theory phenomenon, which often occurs when
a gauge symmetry is spontaneously broken at a high scale.

In superspace, there are two terms that are important for anomaly inflow.
Consider a model with a chiral multiplet $S$,
an abelian vector multiplet $V_X$, and an $SU(N)$ gauge group
with which the $U(1)$
has a mixed ($U(1) SU(N)^2$) anomaly.
The FI term and the mass for the $U(1)$ gauge boson both arise
from the gauge-invariant term
\eqn\KahlerTerm{ L \supset \int d^4 \theta ~K( S + S^\dagger + \delta V_X)
= V_X|_{\theta^4 } K'(s + \bar s) + {1\over 2 } K''(s+\bar s) ( \del \phi + A_X)^2
+\dots, }
where $ \phi = {\rm Im} s$, with $s$ being the lowest component of $S$.
In string models where $U(1)_X$ is an open-string gauge
symmetry, $\phi$ is a mode of a closed string RR field.
Under a $U(1)_X$ gauge transformation,
$ A \mapsto A + d \lambda, \phi \mapsto \phi - \lambda $.
Note that the first term on the RHS of \KahlerTerm\ can be interpreted
as a (dynamical) FI term, while the second term is a field-dependent
gauge boson mass $M_X$.
In type II string models, this mass is moduli-dependent and can
satisfy $M_X \ll M_s$ \IbanezQP.

The other important term for anomaly inflow
(which arises from one-loop gauge-coupling renormalization in the
high energy theory) is
\eqn\Scoupling{ L\supset \int d^2 \theta~ S ~\tr W_\alpha W^\alpha ,}
the contribution of $S$
to the holomorphic gauge coupling of
the nonabelian gauge group, whose gaugino superfield is
$W^\alpha$.  The shift of $\phi$ under the anomaly then cancels the
anomalous variation of the action in an $SU(N)$ instanton background.

Now, one can ask, ``how should $K$ depend on $S$?"
As a simple field-theoretic example, if the axion arises as the phase
of a
chiral superfield $X$ that gets a high-scale vev,
then the FI term comes from the ordinary
canonical gauge invariant kinetic term for $X$:
$$ \int d^4 \theta ~X^\dagger e^{2\delta_X V} X  .$$
In this case, $\vev{ X} = M e^{S} $
and so
$ K \sim e^{ S + S^\dagger + 2 \delta_X V } $.
In string theory, the relevant K\"ahler potentials are more
complicated, but the general structure remains as in
\KahlerTerm.

Note that the massive $U(1)$'s, for any finite $M_X$, should $\it not$ simply
be thrown away.  Their presence can qualitatively alter the vacuum
structure.   One can see this in two ways:

\noindent 1)
Suppose the $U(1)_X$ gauge boson is relatively light compared to the
cutoff $\Lambda_{UV}$ of the effective field theory, $M_X \ll \Lambda_{UV}$.
Then, one should
include the gauge supermultiplet
in the low-energy effective theory, and impose the related
D-term conditions.  These change the potential and alter the vacuum
structure.  In a theory with chiral multiplets $\phi_i$
of charges $q_i$, the full potential will take the form
\eqn\potis{
V = \sum_i \left| {\partial W \over \partial \phi_i} \right|^2
- {1\over 2 g_X^2} D_X^2 - {1\over  g_X^2} D_X \left( \sum_i q_i \vert \phi_i \vert^2 -
\xi \right)}
with $\xi$ the FI term.

\noindent 2)
Suppose instead that one is interested in constructing an effective theory
integrating out the $U(1)$, $\Lambda_{UV} \leq M_X$.
Integrating out the massive $U(1)$ gauge boson
generates a correction to the K\"ahler potential for the chiral multiplets
\ArkaniHamedNU\
\eqn\ADMcorrection{ \delta K = - {g_X^2 \over M_X^2} q_i q_j \phi^{\dagger i}
\phi_i \phi^{\dagger j}\phi_j ~.}
This correction is of dimension greater than four, so one might feel that
it cannot influence the vacuum structure.  However, one can see that at
a stationary point of the full potential $V$ \potis, ${\it gauge ~invariance}$
implies that
\eqn\reln{\langle D_X \rangle = -{g_X^2 \over M_X^2} \sum_i q_i \vert
\langle F_i \rangle \vert^2~}
where
\eqn\xmass{M_X^2 = g_X^2 \sum_i q_i \vert \phi_i \vert^2~.}
In the theory with $U(1)_X$ integrated out, the correction to the effective
potential which arises from plugging nonzero $F_i$ into \ADMcorrection,
then ${\it precisely}$ reproduces the effect of the D-terms from the massive
$U(1)$ (at least as far as scalar soft masses are concerned).

One must be careful when using this logic.  If $M_X$ exceeds
some other threshold where new states are produced (notably $M_s$ in a
generic string compactification), other corrections to $K$ will enter at
the same order and should not be ignored.  Then, there is no consistent
effective theory which only includes the quartic correction \ADMcorrection\
to the canonical K\"ahler potential.  So for this formalism to be strictly
useful, one should work in a regime where \eg\ $M_X \ll M_s$.
$M_X$ should be the first threshold where new physics is encountered.

\subsec{The axion charges in the cone over $dP_1$}

To compute the anomaly-cancelling axion charges in our theory, it is useful
to work in the context of a concrete compactification of the cone.
We will use a construction modeled on the examples of \DiaconescuPC.
\ifig\dcover{The Calabi-Yau singularity.}
{\epsfxsize4.5in\epsfbox{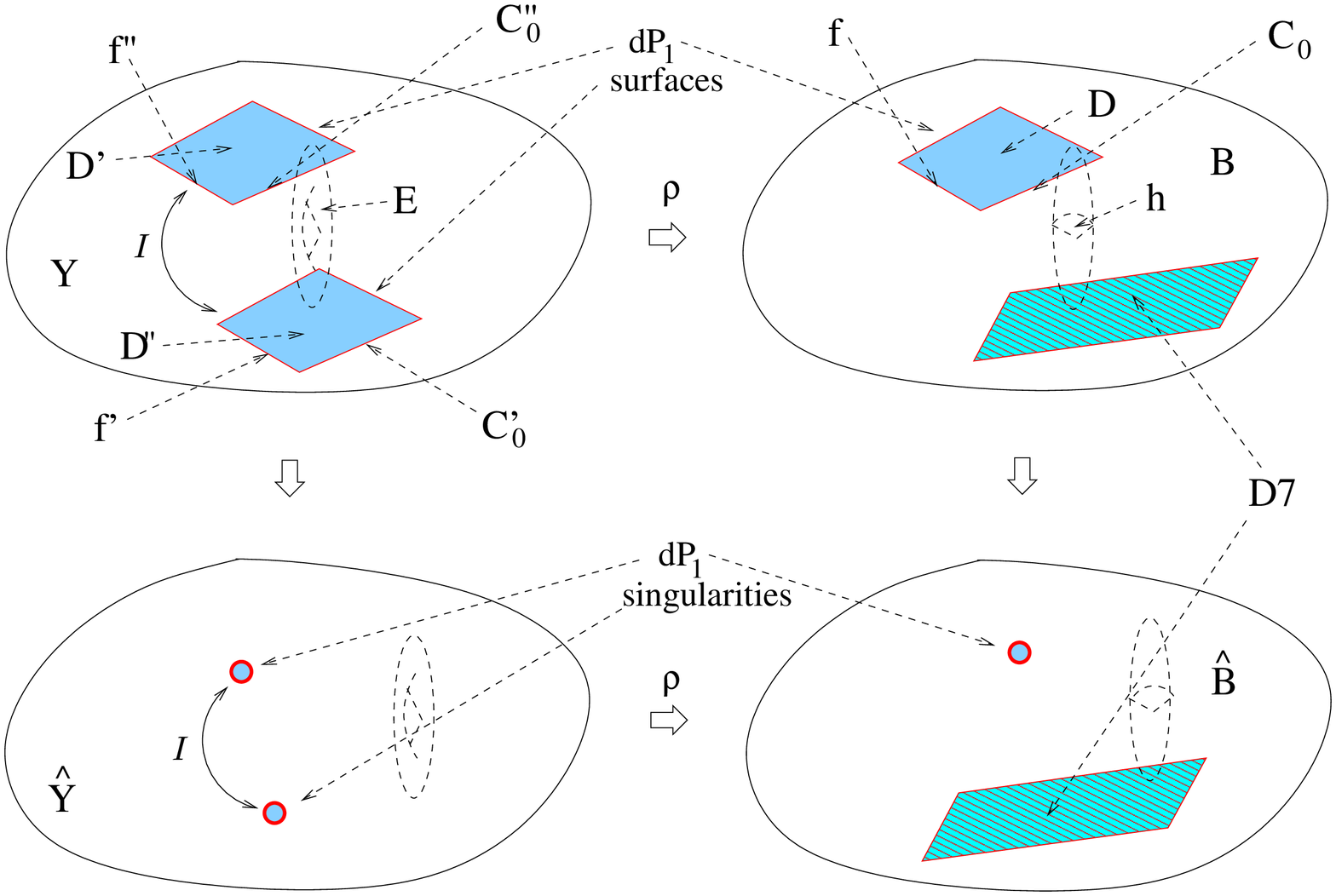}}
We start with the projective threefold $$ B \equiv \IP \left( \CO_{\IF_1}
\oplus \CO_{\IF_1}( \CK_{\IF_1} ) \right). $$ The negative section of $B$ is a
contractible surface $D$ isomorphic to a del Pezzo $dP_1$. Consider now a
Calabi-Yau hypersurface $Y$ given by a quadric in the toric fourfold
$Z=\IP \left( \CO_{B} \oplus \CO_{B}( \CK_{B} ) \right).$ $Y$ is a torus fibration
over $dP_1$ with two sections and has $h^{1,1}(Y)=4,~h^{2,1}(Y)=148$. There exists a
$\IZ_2$ involution $\CI:Y\rightarrow Y$
such that the quotient $Y/\CI$ is isomorphic to $B$ and the inverse image of $D$ under
the map $\rho:Y\rightarrow B$ consists of a pair of surfaces $D',D''$ isomorphic to
$dP_1$ which are contractible inside $Y$. The resulting singularities admit first
order deformations which are obstructed at second order.

By construction, the local neighborhood of $D$ in $B$ is Calabi-Yau, so we can use facts
about the quiver gauge theory that were derived by studying singular Calabi-Yau threefolds.
In a more general compactification of F-theory on a fourfold, the local neighborhood of
the singularity in $B$ would not necessarily be Calabi-Yau, and the results could differ.
\vfill\eject

\subsubsec{The surviving axions}

Before describing the geometry in more detail, we pause here to explain
some differences between these models and more commonly studied D3/D7
orientifolds.  One important fact is that because the involution exchanges
two copies of $dP_1$ (call them $D$ and $D^\prime$), it also exchanges
the curves $C_0, f \subset D$ with $C_0^\prime, f^\prime \subset D^\prime$.
In the theory
with {\it no} orientifold quotient, one would have obtained (from this part of
the geometry) the six RR axions\foot{We work in the normalization that RR forms
shift by  integral closed forms  under global gauge transformations.}
$$\phi_{C_0} = \int_{C_0} C_2,~~\phi_{C_0^\prime} = \int_{C_0^\prime} C_2$$
$$\phi_{f} = \int_f C_2,~~\phi_{f^\prime} = \int_{f^\prime} C_2$$
$$\phi_{\IF_1} = \int_{\IF_1} C_4,~~\phi_{\IF_1^\prime} = \int_{\IF_1^\prime}
C_4~.$$

The orientifold involution acts by ${\cal I}$ composed with
$\Omega (-1)^{F_L}$ where $\Omega$ is worldsheet parity and $F_L$ is
left-moving fermion number.  Normally (see \eg\ p.6 of \KST), the action
of $\Omega (-1)^{F_L}$ on $C_2, C_4$ is $-{\rm 1}, +{\rm 1}$, and one
projects out the $C_2$ axions.  However, here, by taking appropriate
combinations from the two copies of the shrinking del Pezzo, one gets both
$C_2$ and $C_4$ axions.  Namely, the combinations
\eqn\hminusaxions{
\eqalign{ \tilde \phi_{C_0} &= \phi_{C_0} - \phi_{C_0^\prime}\cr
\tilde \phi_f &= \phi_f - \phi_{f^\prime} \cr
\tilde \phi_{\IF_1} &= \phi_{\IF_1} + \phi_{\IF_1^\prime}}}
all survive the orientifold projection.
Henceforth, to simplify life, we will abuse notation and
call these surviving combinations
$\phi_{C_0}$, $\phi_{f}$ and $\phi_{\IF_1}$.


\subsubsec{The geometry of $B$}

$B$ can be described torically (or in the language of the gauged linear sigma
model (GLSM) \WittenGLSM) as follows.
There are six chiral multiplets in the GLSM, charged
under three $U(1)$ gauge symmetries:
\eqn\glsmcharges{
\matrix{& D_1 & D_2 & D_3 & D_4 & D_5 & D_6 \cr \cr
             &  1      &0    & 1       &   0      &   -2     &    0 \cr\cr
             &  0      &1    & -1      &   1      &   -1     &    0 \cr\cr
             &  0      &0    & 0       &   0      &   1      &    1
             .}}
In terms of the geometry of the $dP_1$ described earlier, one should think of
$D_2$($D_3$) as the $\IP^1$-fibration over $f$ ($C_0$).
$D_5$ is the $\IF_1$ base.

It follows from the toric data that there are three linear
equivalences:
$$D_2 = D_4,~ D_6 = D_5 + 2 D_3 + 3 D_2, ~D_1 = D_2 + D_3 .$$
The nonzero triple intersections are
$$ D_5^3= 8, ~D_5 D_2 D_3 =1 , ~D_5^2 D_2 = - 2, ~
D_5^2 D_3 = D_5 D_3^2 = -1. $$
A consequence of the fact that a neighborhood
of the del Pezzo is locally CY is the relation
\eqn\cyrelation{ D_5^2 = - 2 D_3 D_5 - 3 D_2 D_5 .}
It says that the self-intersection of the base
is its own canonical curve.

Note from \glsmcharges\ that the geometry $B$ has an $SU(2)$ symmetry
under which the coordinates $D_2, D_4$ transform as a doublet.
Any particular representative of the linear equivalence class of
$D_2$ breaks this symmetry.
This symmetry is the geometric origin of the $SU(2)_F$ flavor symmetry
of the field theory described in \S2.

In order to compute the axion charges, we need to evaluate the
topological terms in the brane worldvolume action.  Recall that for a
brane characterized by some sheaf $V_{\alpha}$ on $B$, these couplings
are summarized by (in the units $2\pi \alpha'=1$)
\eqn\topterm{
{i\over 2} \int_B \sum_{p~ {\rm even}} C^{(p)} \wedge
\sqrt{\hat A(B)} \wedge {\rm ch} V_{\alpha} \wedge {\rm tr_{\alpha}} F
\wedge F ~.}
Therefore, we will need  the Chern classes of $B$ and
the bundles characterizing the quiver nodes.

The total Chern class of $B$ is
$$ c(B) = \prod_{i=1}^6 ( 1 + D_i ) = ( 1 + D_2 + D_3)(1+ D_2)^2 ( 1 +  D_5) ( 1 + D_5 + 2D_3
+ 3D_2) ( 1 + D_5) .$$
It follows that the Todd class and A-roof genus of $B$ are
$$ \td(B) = 1 + {c_1\over 2 } + { c_1^2 + c_2 \over 12 } + \dots
= 1 + D_5 + 2 D_3 + 3 D_2 + \dots $$
$$\hat A(B)=e^{-{c_1(B)\over 2}}\, \td(B).$$

\subsubsec{Pushing forward our branes}

In order to compute the Chern-Simons
terms in the worldvolume Lagrangians, we will pull back the
quiver branes
to bundles on $B$.
To do this we apply the Grothendieck-Riemann-Roch${}^{\rm TM}$
formula to the inclusion map
$ i: D_5 \hookrightarrow B $.
Applied to $\CL_1 = \CO_{\IF_1}$, this says
\eqn\GRRRR{i_\star ( \ch \CL_1 \td(D_5) ) = \ch( i_\star \CL_1) \td(B). }
Using $\td(D_5 ) = 1 + C_0 + {3\over 2} f + [{\rm pt}_{\IF_1}]$, $\ch(\CL_1) = 1$,
and
$$ i_{\star} (  \ch \CL_1 \td(D_5) ) = D_5 + D_5 D_3 + {3\over 2} D_5 D_2 +
[{\rm pt} ] $$
we get
$$  \ch( i_\star \CL_1) = D_5 - D_5 ^2 -  D_3 D_5 - { 3 \over 2 } D_2 D_5 + \dots
= D_5 + D_3D_5 +{3 \over 2} D_2D_5 +\dots.$$
Throughout, we are dropping ``6-form'' terms in the computation that
would only
determine the charges of the 10d RR axion $C_0$ -- it follows from
\IbanezQP\ that $C_0$ does not shift in the relevant Green-Schwarz mechanism.

Similarly, we find
$$  \ch( i_\star \CL_2) = D_5 - D_5 ^2 - \half D_2 D_5 + \dots
= D_5 + 2 D_3D_5 +{5\over 2} D_2D_5 + \dots$$
$$  \ch( i_\star \CL_3) =- D_5 + D_5 ^2 + D_3 D_5 + \half D_2 D_5 + \dots
= -D_5 - D_3D_5 -{5 \over 2} D_2D_5 +\dots$$

\subsubsec{Axion charges}

Finally, we can identify the charges we should assign to various RR
axions, to cancel the worldvolume anomalies.
The worldvolume theory of a D-brane on $B$
with gauge sheaf $V_\alpha$
contains the coupling \topterm.
Note that we are assuming that the gauge bundle factorizes
between $B$ and the four noncompact directions.
In keeping with our earlier definitions, we let
$$\phi_f = \int_B C^{(2)} \wedge D_5 \wedge D_2, ~~
\phi_{C_0} = \int_{B} C^{(2)} \wedge D_5 \wedge D_3,~~ \phi_{\IF_1}
= \int _B C^{(4)}\wedge D_5= \int_{\IF_1} C^{(4)} ;$$ we are using a
notation where $ \int _B D \wedge \eta = \int_D \eta ~ (\forall
\eta) $, so $D$ is a two-form. Also, let us denote
$$\Phi=\phi_{\IF_1}+\phi_{C_0}+{3\over 2}\phi_f.$$
The anomalous variation of the worldvolume fermion measure (which
captures the mixed anomalies of $U(1)_{1,2,3}$ with the $SU(M),
SU(2M)$ and $SU(3M)$ factors, respectively)\foot{ The same choice of
local counterterms will also cancel variations from mixed
$U(1)_{a}U(1)_bU(1)_c$ ($a,b,c\in \{1,2,3\}$) anomalies.} is
$$
\eqalign{ \delta \Gamma ={i\over 2}\Biggl(-
\tr_1 F\wedge F [ & + 0 \lambda_1  -6M \lambda_2  +6M \lambda_3 ] -\cr
\tr_2 F\wedge F [ & +3M \lambda_1  + 0 \lambda_2 - 3M \lambda_3 ] -\cr
\tr_3 F\wedge F [    & -2M \lambda_1  +2M \lambda_2 + 0 \lambda_3 ]\Bigr) ,}
$$
where $\lambda_{1,2,3}$ are gauge parameters for $U(1)_{1,2,3}$.
We set this equal to minus the variation of
$$ \eqalign{S_{\rm CS} ={i\over 2}\Biggl(
\tr_1 F\wedge F [ & \Phi + { 0 } \phi_f +  0 \phi_{C_0} ] + \cr
\tr_2 F\wedge F [ & \Phi +  \phi_f +   \phi_{C_0} ] + \cr \tr_3
F\wedge F [ & -\Phi -  \phi_f  +  { 0} \phi_{C_0} ] \Biggr)}
$$
(note that the $D_5^2$ term in the chern characters contributes
through the relation \cyrelation). We find, in multiples of $M$, the
following charges \eqn\AxionCharges{ \matrix{ & U(1)_1 & U(1)_2 &
U(1)_3 \cr\cr e^{i\phi_{C_0}} & 1 & 2 & -3  \cr\cr e^{i\phi_f} & 2 &
4 & -6  \cr\cr e^{i\Phi} & 0 & -6 & 6. }}

\medskip
A check on our algebra is the fact that
no field transforms under the
overall decoupled $U(1)$, which has $\lambda_1 = \lambda_2 = \lambda_3$
-- \ie, the sum
of the entries in each row vanishes.
Note that the combination
$ 2 \phi_{C_0} - \phi_f $ is neutral;
$ 2 C_0 - f $ is the cohomology class of the
net D5 brane charge of the fractional brane.

\newsec{Strings Between the Instanton and the Quiver}

To determine whether a Euclidean D3 contributes to the superpotential,
we will need to know
about the topology of the cycle it wraps in the Calabi-Yau, and we will
need to determine its spectrum of ``Ganor strings'' -- strings which
stretch from the Euclidean brane to the (occupied) nodes of the quiver.
Such strings act as collective coordinates of the instanton.

In our situation, both the Euclidean D3's and the quiver nodes wrap
a surface $S \subset B$ (the $\IF_1$), and carry different
line bundles $\CL_A$, $\CL_B$ over $S$.
The correct way to work out the spectrum is then to do a twisted
reduction of the $D3-D7$ hypermultiplet on $S$.

\subsubsec{Twisted reduction}

Here we discuss the twisted reduction of the hypermultiplet
of 3-7 strings, following
a similar analysis for M5-brane fields in \WittenBN.

The normal bundle to a del Pezzo surface $S$
decomposes as $ N \oplus T\IR^4$, where
$N$ is the normal bundle to $S$ in $B$.
Because we are studying a situation where
the neighborhood of the surface $S$ is Calabi-Yau,
$N= K$, where $K$ is the canonical bundle of $S$.
Following \WittenBN, we can identify
the rotations of this normal bundle to the divisor with
an R-symmetry of the field theory.  This is because
the superspace coordinates of the instanton
arise from 3-3 strings; these transform as sections of
the spin bundle derived from $N$,
$S^\prime \equiv N^{1/2}_{1} \oplus N^{-1/2}_{-1} $,
where the subscripts denote R-charges.

In the neighborhood of the del Pezzo singularity,
the ten dimensional Lorentz group is broken to
$$ SO(4)_{\IR^4} \times SO(4)_{dP_1} \times SO(2)_{\perp} \subset SO(10).$$
Let $S^{\prime\prime}_\pm $ be the (trivial) chiral spin bundles on $\IR^4$.
The spin bundle\foot{
Note that while $dP_1$ is not spin,
implying that there is a Freed-Witten anomaly for
branes wrapping it,
it is spin${}^{\rm c}$.
Both the D7-brane gauge bundle
and the D3-brane gauge bundle
are twisted by $K^{1/2}$ \Mooreii .  Since we
will be interested in the dynamics of strings
stretching between them,
these factors cancel and this anomaly
does not affect our analysis.}
on $dP_1$
decomposes into chiral parts as $S_+ \oplus S_-$ where
\eqn\splusdP{ S_+ = K^{1/2} \oplus \left( K^{1/2} \otimes \Omega^{0,2} \right) }
\eqn\sminusdP{ S_- = K^{1/2} \otimes \Omega^{0,1} .}

Using the facts about $\#_{ND}=4$ systems (page 162 of \Polch), we find that
the fermions transform as sections of
\eqn\fermibundle{
\left( S^\prime \otimes S_+\otimes
\CL_A \otimes \CL_B^\star \right)
\oplus
\left(S^\prime  \otimes S_-\otimes
\CL_A^\star \otimes \CL_B \right). }
Using $N=K$, this is
\eqn\reducefermibundle{
\eqalign{
& \CL_A \otimes \CL_B^\star \otimes
\left( K_{1} \oplus
(K_{1} \otimes \Omega^{0,2}) \oplus \CO_{-1} \oplus \Omega^{0,2}_{-1}
\right)
\cr
 \oplus ~~&
\CL_A^\star \otimes \CL_B \otimes \left( ( K_{1} \otimes \Omega^{0,1}  ) \oplus
\Omega^{0,1}_{-1} \right) .}
}
The dimensions of these summands, labelled by
4d chirality and R-charge, are:
\eqn\dims{
\eqalign{
&\left(
h^{0,2}(\CL_A^\star \otimes \CL_B)_{1} + h^{0,0}(\CL_A^\star \otimes \CL_B)_{1}
+ h^{0,0}(\CL_A \otimes \CL_B^\star)_{-1} + h^{0,2}(\CL_A \otimes \CL_B^\star)_{-1} \right) \cr
&+ \left( h^{0,1}(\CL_A \otimes \CL_B^\star)_{1} + h^{0,1}(\CL_A^\star \otimes \CL_B)_{-1}
\right). }}
In deriving \dims\ we have used Serre duality:
\eqn\serre{ \left(H^0( \Omega^{0,p} \otimes K \otimes L ) \right)^\star
= H^0( \Omega^{0, 2-p} \otimes L^\star ) .}
The bosons transform as sections of
$\left( S^{\prime\prime}_+ \otimes \CL_A \otimes \CL_B^\star \right)
\oplus \left( S^{\prime\prime}_-\otimes \CL_A^\star \otimes \CL_B \right)$
(recall that a hypermultiplet
contains an $SU(2)$ doublet of complex bosons);
the trivial spin bundles $S^{\prime\prime}_{\pm}$ over $\IR^4$
simply give the bosons multiplicity two.


In the following we have weighted the spectrum by
signs in order to omit modes which can pair up in a manner consistent with
gauge invariance and R-charge conservation.
The end result is that the spectrum consists of:

\noindent
$\bullet$ A net number of bosons given by
\eqn\bosspec{
n_{bos}(\CL_A,\CL_B)=2\bigl(h^{0}(S,\CL_A \otimes \CL_B^*) - h^{0}(S,\CL_B \otimes \CL_A^*)~\bigr).}

\noindent
$\bullet$ A net number of fermions given by
\eqn\fermspec{n_{ferm}(\CL_A,\CL_B)=\chi(\CL_A \otimes \CL_B^*)-
\chi(\CL_A^* \otimes \CL_B)}
where
$$\chi(\CL_A \otimes \CL_B^*) = \sum_{i=1}^{3} (-1)^p h^{p}(S,\CL_A \otimes
\CL_B^*)~.$$


\subsec{Ganor strings for a general class of bundles}

Next we apply the results of the previous subsection to
compute the spectra of Ganor strings
for general instantons in the $dP_1$ example.
These will be well summarized by extended
quiver diagrams, where one includes a node for the instanton
appended to the regular quiver, and adds arrows
for the strings which stretch between the instanton and the quiver nodes.
In general, one can then read off gauge invariant couplings between the
instanton collective coordinates and the quiver fields, which will be
relevant when evaluating the integral over the instanton modes (which is
really an ordinary integral, not a functional integral).

The most general line bundle of interest for us is
$X_{ab} = {\overline{ {\cal O}_{D_5}(aC_0 + bf)}}$.
The D3-instantons wrapping divisor in a  class $[D_5+cD_3+dD_2]$
for non-zero $c$ and/or $d$ intersect ``divisor at infinity" $D_6.$
Therefore, such instantons have more zero modes and
are likely to vanish;
the detailed analysis of their contribution is model-dependent.
We now compute the cohomology groups using \eg\ \fulton,
\eqn\hzero{
h^0(dP_1, \CO( nC_0+ m f) ) = \#\{ (x,y) \in \IZ^2 | x \geq -m, y \geq 0, -x-y \geq 0, -y \geq -n \}
}
and Euler number of the bundle on $D_5$ using the GRR formula
\eqn\GRRagain{ \chi(X) = \int_{D_5} \ch(X) \wedge \td(D_5). }
 We find that
the number of fermionic zero modes between $X_{ab}$ and $\CL_{1,2,3}$ is
\eqn\zerom{n_{ferm}(X_{ab},\CL_{1,2,3}) = (a+2b,~-3+a+2b,~2-a-2b) ~.}
Note that the answer depends only on $a+2b$ which we from now on set equal to 2
so that the instantons have  $U(1)_B$ charges $(2,-2,0).$ This choice is
motivated by  the fact that operators of greatest interest need those charges cancelled.

The number of bosonic zero modes depends on b. There are no bosonic zero modes
for $b>1$ and $b\le -1,$ meanwhile for $b=1$ we find
\eqn\zeromii{n_{bos}(X_{01},\CL_1)=4,\quad
n_{bos}(X_{01},\CL_2)=-2,\quad n_{bos}(X_{01},\CL_3)=0.}

\newsec{Instanton Effects}

Equipped with the results of \S4, we can now try to classify instantons
by considering arbitrary line bundles on $S$ and integrating out their
Ganor strings.
The contribution of each such instanton
is proportional to an exponential of RR axions,
and therefore carries a definite set of $U(1)_B$ gauge charges
by the results of \S3.2.  One check on our results should be that each
non-vanishing contribution enjoys a cancellation of $U(1)_B$ charges between
the instanton action and the field theory operator arising from the integral
over Ganor strings.

\subsec{Which topologies can contribute?}

Before we consider explicit examples, we
would like to discuss which instantons can contribute to the superpotential.
Let us first recall the constraints on topology of the instantons
given in \WittenBN.  Using the
duality between F-theory and M-theory, each Euclidean D3 instanton can
be related to an M5 brane instanton in the F-theory fourfold $X$, wrapping
some divisor $D_M$.
In \WittenBN, Witten argued that, in the absence of the background fluxes,
the necessary condition for D3-instantons to contribute is $\chi(D_M)=1$, where
\eqn\rcont{\chi(D_M) = \sum_{p=1}^{3} (-1)^p h^{0,p}(D_M)~.}
$\chi(D_M)=1$ is necessary but not sufficient
for instanton to generate W since it counts 3-3 fermion zero modes with
sign. The sufficient condition is, of course, to saturate all fermion zero modes.

As further shown in \VP, in the presence of gauge flux $F$ on
the world-volume of a D3-brane, the  Dirac operator acting
on world-volume fermions is deformed (see eq.\ 69 of \VP).
This allows a possibility to lift some or (in the lucky case) all
of the non-universal 3-3 fermion zero modes, i.e. modes which
come, in the M-theory language, from $h^{0,p}(D_M)$ for $p>1.$

There is also another mechanism of lifting(in some cases) non-universal 3-3 fermion zero modes, which is due to background fluxes \Saul,\Kalosh,\Kaloshii.

When D3-instanton intersects space-filling branes,
the condition for the instanton
to contribute is modified since 3-7 and 7-3 zero modes, discussed in \S4,
should be saturated as well.
In general, integrating out these 3-7 and 7-3 zero modes leads to
some operator
${\cal O}_D$ in the quiver gauge theory to arise as a prefactor of
exponent of instanton action.
The subscript
is a reminder that the precise operator is determined by the integral over
the instanton collective coordinates; it is a function of the instanton.


\subsec{BPS instantons}

One could worry that a semiclassical description involving
branes wrapping a weakly curved space
might break down near the quiver locus in K\"ahler moduli space
where D7 branes and anti-D7 branes are simultaneously BPS.
However, we are computing corrections to the superpotential,
which is a holomorphic function of chiral superfields.
Since the partners of K\"ahler moduli are axions,
it is natural to expect that all of the dependence of the
superpotential on the K\"ahler moduli
appears in exponentials.
We will make this assumption.


At the quiver locus in K\"ahler moduli space, where the quiver gauge theory
exists, the central charges of the four sheaves on $\IF_1$
$$\CO_{\IF_1},\quad \CO_{\IF_1}(h),\quad \overline{\CO_{\IF_1}(f)},\quad
\overline{\CO_{\IF_1}(C_0)}$$ are aligned, \ie,
have the same phase. The central charge $Z(X_{ab})$ of
 D3-instanton with bundle
$X_{ab}$ can be expressed  as linear combination:
\eqn\lincomb{Z(X_{ab})={1\over 2}(a^2-2ab+a+2b-2)Z(\CO_{\IF_1})
+{1\over 2}(a^2-a-2ab)Z(\CO_{\IF_1}(h))}
$$+{1\over 2}(a^2-a+2b-2ab)Z(\overline{\CO_{\IF_1}(f)})
+{1\over 2}(a^2-2ab+a)Z(\overline{\CO_{\IF_1}(C_0)}).
$$
For $a+2b=2$ all coefficients in the linear combination \lincomb\
are non-negative. Therefore, all of the instantons $X_{ab}$ are aligned
with nodes of the quiver. This ensures that such instantons are BPS
and 3-3 strings have two universal fermion zero modes $\theta^{\a}$ coming from
breaking half of ${\cal N}=1$ supersymmetry in $\IR^4.$

Please note that in the absence of gauge flux on D3-instanton,
there are also two non-universal zero modes $2h^{(0,2)}(K,D_5)=2.$
These are lifted by the world-volume gauge flux $[F]=aC_0+bf$
as follows from the analysis of deformed world-volume
Dirac operator \VP.

Finally, there are 3-7 and 7-3 zero modes discussed in \S4 which we integrate out in the next section.
\subsec{First example: The ADS instanton as a D-brane}

Consider our quiver for $M=1$.  The $SU(3)$ node has two flavors of quarks,
and should generate a superpotential via the Affleck-Dine-Seiberg
instanton effect \ADS.
(It is important that precisely for $N_f = N_c-1$, the dynamical superpotential
in SUSY QCD is actually an instanton effect, and not a result of other
strong dynamics as it is for $N_f \leq N_c-2$).
A D3 brane wrapping $\IF_1$ with the ${\it same}$ bundle as $\CL_3$ has
precisely the correct action to be interpreted as the gauge theory
instanton.
This is the special case of \zerom\ with $a=0, b=1$.
Can we compute this contribution to the superpotential
by quantizing its Ganor strings? (The reader
is advised to see also \Bershadsky\ for a telegraphic
discussion of a closely-related question.)

The appropriate extended quiver diagram, including a node for the
instanton and the Ganor strings, appears below:
\ifig\extquivertwo{The extended quiver for the $SU(3)$ instanton $X_{ADS}\equiv X_{01}.$}
{\epsfxsize1.5in\epsfbox{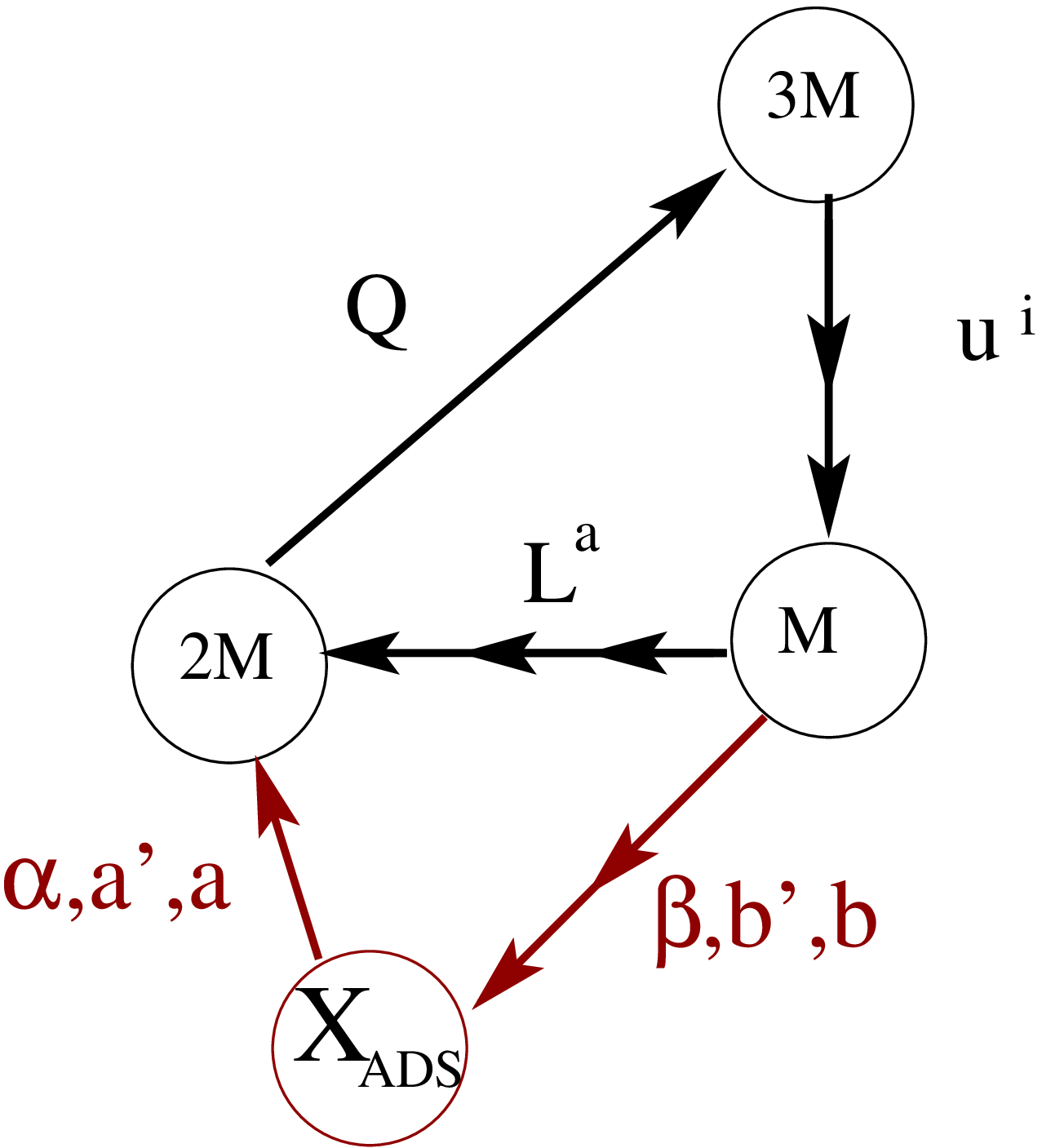}}

\noindent
This case is special in that $X_{01}\otimes\CL_\alpha$ have holomorphic
sections; the analysis of the previous section then implies
that there are {\it bosonic} zeromodes $a,b_j,a', b'_j$,
in addition to the fermions $\alpha, \beta_j$.
The gauge charges are as indicated by the arrows in \extquivertwo.
A gauge invariant coupling between the quiver fields and the collective
coordinates, which should be generated at the disc
level, is
\eqn\discamp{\tilde L = \alpha Q \cdot u^i M_{i}^{j} \beta_j
 +a Q \cdot u^i N_i^j b_j+a' Q \cdot u^i N_i^{'j} b'_j~}
where $M, N,N'$ are some flavor matrices.

The contribution of the instanton $X_{01}$ is proportional to
$$e^{-S_{\rm instanton}}=e^{-S_1+\theta^{\a}\psi_{\a}+{\tilde L}}$$
where \eqn\defsi{ S_1:=t_{\IF_1}+t_f,\quad t_{\IF_1}={1\over 4\pi
g_s}\int_{B}J^2D_5+i\Phi,\quad t_f={1\over 2\pi g_s}\int_{B}
JD_2D_5+i\phi_f.} $\theta^{\a}$ are two fermion zeromodes of 3-3
strings and $\psi_{\a}$ are fermion superpartners of $S_1.$ Volumes
are measured in string units $\alpha^\prime={1\over 2\pi}$ and as in
section 3 RR fields are normalized to have integral periodicity.

If we parametrize K\"ahler form $J$ as
$$ J=r_5D_5+r_2D_2+r_3D_3$$
we find
$${\rm Re}(S_1)=\half(8r_5^2-r_3^2-2r_3r_5-4r_2r_5+2r_2r_3)+r_3-2r_5.$$

To find the superpotential we compute two-point correlator of
fermions $\psi_{\a}$ in the instanton background:
$$\langle \psi_1^{\dg} \psi_2^{\dg} \rangle_{instanton}=
G_{\psi_1}G_{\psi_2}\p_{S_1}\p_{S_1} W$$
 where $G_{\psi}=\langle \psi^{\dg}\psi \rangle $.
Integral over 3-3 fermion zero modes pulls down $\psi_1\psi_2$
and we are left with integral over the 3-7 zero modes
$a,b,a',b',\alpha,\beta$:
\eqn\wequation{
\p_{S_1}\p_{S_1} W=e^{-S_1}\, \int da ~d^2b ~da'~d^2b'~d\alpha ~d^2\beta~
e^{\tilde L} \sim
{ \det M \det_{fj} (Q^f \cdot u_j) \over
\det N \,\det N'\, \det^2_{fj} (Q^f\cdot u_j)}}

\noindent
From \wequation\ we immediately find
the following contribution to the superpotential\foot{$\Lambda\equiv \Lambda_3.$}
\eqn\adscontrib{W_{ADS}={\Lambda^7 \over Z}e^{-S_1}}
where $Z=\det_{fj} (Q^f\cdot u_j)$ and $\Lambda^7$ comes from
path integral normalization.

Note that:

\noindent
$\bullet$ The factor of two in the multiplicity of bosons \bosspec\
was crucial for obtaining the ADS result.


\noindent
$\bullet$ The $U(1)$ gauge charges of the instanton action (inferred
from the axion couplings) nontrivially cancel with those of ${\cal O}$
to yield a gauge invariant operator.

\noindent
$\bullet$ Other couplings besides \discamp\
are conceivable.
For example, the full disc amplitude could give
$\tilde L  = f\left(\alpha Q  u \beta\right) $
where $f$ is some more general function than $f(x)=x$.
The important point is that it can only depend on these combinations of
fields, and changing variables in the $\alpha, \beta$ integrand
by \eg\ $ \alpha' \equiv \alpha Q \cdot u, \beta' \equiv \beta$
gives the expression above, times some integral does not
depend on the quiver fields.
This remark applies in all the cases we will discuss.

\subsec{Second example: A ``stringy'' deformation of the field theory
superpotential}

Now, consider the case $b>1$ or $b\le -1$.  The relevant extended quiver
diagram appears below:
\ifig\extquiverone{The extended quiver for the
$X_{ab}$ instantons, when
$2 = a+2b$ and $b>1$ or $b\le -1$.
$\alpha,\beta$ are fermionic.}
{\epsfxsize2.0in\epsfbox{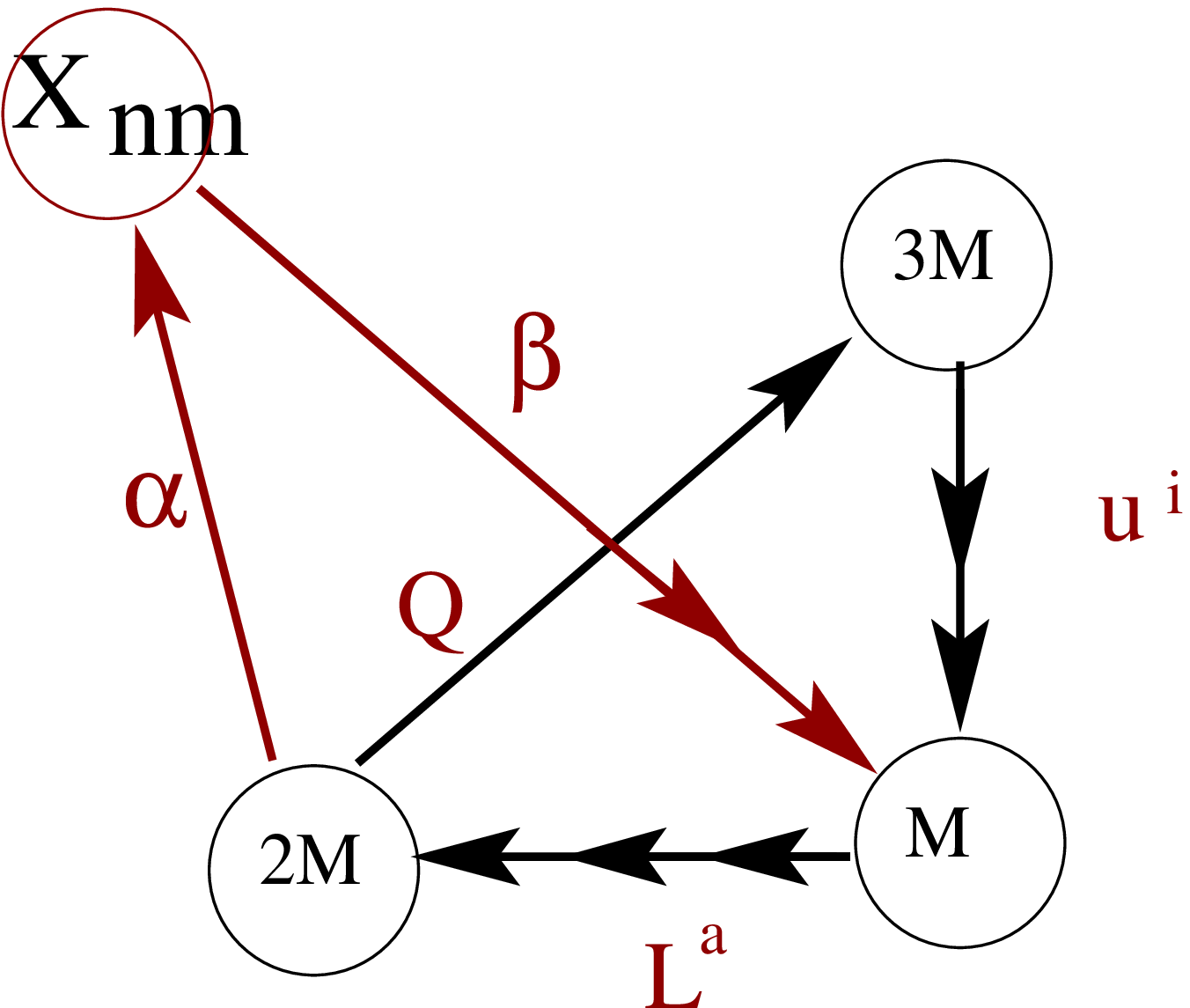}}

\noindent
A gauge invariant coupling which is expected to arise from a disc amplitude
is
\eqn\ltwo{L = \alpha L^a M_{aj}\beta^j~}
where $M$ is some flavor matrix
which breaks the $SU(2)_F$
and the $SU(2)$ rotating the $\beta^j$ down
to a diagonal subgroup.
It is worth noting that the $SU(2)$ which rotates the fermion zeromodes
has the same origin as the $SU(2)_F$.  It arises from
the action of the $SU(2)$ isometry of the $dP_1$ on its
cohomology.  The two classes generating the $\beta^j$
arise from the $SU(2)$ doublet of sections $z^j$ of $\CO(1)$ over the sphere
acted on by this $SU(2)$.  It is therefore not at all surprising to
find that this geometric coupling preserves only one $SU(2)_F$.

Integrating over the fermions
$\alpha,\beta$ generates
an operator of the form $\sum_a c_a {\cal O}_a$, where
${\cal O}_a \sim L^b L^c \epsilon_{abc}$.
What are the values of the coefficients $c_a$?  In the field theory \chargesum,
$L_{1,2}$ form
an $SU(2)_F$ doublet while $L_3$ is an $SU(2)_F$ singlet.
Since the toric 3-fold $B$ has $SU(2)_F$ as a global symmetry,
we have $c_1=c_2=0.$ Including the dependence on K\"ahler moduli, the ``stringy"
instanton correction to $W$ has the form\foot{This
corrects statements made in \stringstalk.}$^{,}$\foot{The
contribution of $X_{20}$ to the superpotential is zero
because of unpaired bosonic zeromodes.
%
}
\eqn\stringycor{
W_{\rm stringy}={\Lambda^7\over M_s^6}V_3\, \sum_{b>1 \& b\le -1}
f(b) \, e^{-S_1+(b-1)S_2}.}
$S_1$ is defined in \defsi\ and $S_2$ is given by
\eqn\defsii{
S_2:=t_f-2t_{C_0},\quad
t_f=\int_{B}({1\over 2\pi g_s}J+iC_2)D_2D_5, \quad t_{C_0}=
\int_B ({1\over 2\pi g_s}J+iC_2)D_3D_5.}
If we parametrize the K\"ahler form $J$ as before
$$ J=r_5D_5+r_2D_2+r_3D_3$$
we find
$${\rm Re}(S_2)=3r_3-2r_2.$$
We do not know the function $f(b)$, but we conjecture that it
 is such that $\sum_{b>1 \& b\le -1} f(b)$
is convergent. $W_{\rm stringy}$ is valid near the quiver locus
$\vert Re(S_2)\vert \ll 1$ so that $\sum_b f(b) < 1$ ensures convergence
of the sum over $b$ in $W_{\rm stringy}.$



The contribution $W_{\rm stringy}$ vanishes in the
limit $M_s \to \infty$ with $\Lambda$ fixed,
which gives the theory studied by \IntriligatorAW.


\subsec{Contributions of D-string instantons}

The curve $C_0$ is isolated.  Hence a Euclidean D-string
wrapping it is a reasonable candidate to
generate interesting instanton effects \WittenEG.\foot{In many D3/D7 models,
there are ${\it no}$ possible contributions from D-string instantons,
since all modes of $C_2$ on the internal space are projected out.  In our
model, some of the surviving axions arise from $C_2$, which is why this
possibility even arises.}
In the next few paragraphs we explain why this
contributes a certain form of zero.
Other curves in the del Pezzo are not isolated,
and therefore are not expected to contribute to the superpotential.

The charge of $\phi_{C_0}$ is
$ (1,2,-3) \times M $.
This is also the number of strings
stretching from a D1 on $C_0$ to branes 1,2,3.
This is
true by the  following calculation.
Let
$$ I= i_\star j_\star \CO_{C_0} $$
($j : C_0 \to \IF_1$ is the inclusion)
be the lift of $C_0$ to $B$.
Then
$$ \int_B \ch(I) \wedge \ch (i_\star \CL_\alpha) \wedge \td (B)
= (1,1,-1)_\alpha $$
counts this number of strings per unit rank.
Again these strings account for the anomalous charge
of the instanton factor.

But what operator can soak up this charge?
We find a satisfying answer as follows.
Consider a euclidean D-string wrapping $C_0$
in the presence of the quiver.
There are massless fermionic strings
in the following representations:
\eqn\FermionCharges{
\matrix{ & U(M) & U(2M) & U(3M) \cr\cr
\psi & 1 & 1 & 3M  \cr\cr
\eta & 1 & \overline{2M} & 1  \cr\cr
\lambda & \overline{M} & 1 & 1.  }}
The $\psi$ is in the antifundamental because
$\CL_3$ is an antibrane.

The following object is gauge invariant:
\eqn\discamplitude{L=  \psi \left( u \lambda +  u\cdot L  \eta \right).}
The dot denotes a sum over $U(M)$ colors.
It is again reasonable to expect that $L$ is generated as
the action for these fermionic modes by disc amplitudes.
There is an important question regarding the flavor structure
of $L$.  Since the instanton on $C_0$ preserves the $SU(2)$
isometry of $\IF_1$, the action $L$ will respect the flavor
$SU(2)_F$ preserved by $W_{\rm tree}$.
This means only $L_{1,2}$ appear.

The integral over these modes of the instanton equals
\eqn\BV{
 A = \int d\psi d\eta d\lambda~ e^L  =
\epsilon^{fg} \epsilon_{ab} \epsilon_{ABC} u_A^i  (u_B^j\cdot L_f^a ) (u_C^k \cdot L_g^b) C_{ijk}^\CX .
}
Here $C_{ijk}^\CX $ is a intertwiner that projects the product of three $\bf 2 $'s onto
the completely antisymmetric $\bf 4$, $\CX=1,\ldots,4$.
But for $M=1$ this is
$$ A = \alpha^a_\CX V_a  \tilde B^\CX, $$
where $V$ was defined above and
$$ \tilde B^\CX \equiv C_{ijk}^\CX \epsilon_{ABC} u_A^i  u_B^j  u_C^k .$$
If the $u$'s are treated as bosonic objects,
this is zero.
So, a gauge invariant superpotential is
$$ e^ {-t_{C_0} } V_a \tilde B^\CX \alpha^a_\CX$$
which is classically zero.
For $M>1$, the contribution also vanishes by
Bose statistics of $u$'s.

\newsec{Vacuum Structure:
On the Capture of the Runaway Quiver}

One of the motivations for this work
was the question of whether the
runaway  of
the $dP_1$ quiver gauge theory \IntriligatorAW\
is cured by embedding into string theory.
Here we apply the results of the previous
sections to address this in the compact model
introduced in \S3.2.\foot{In the discussion below we ignore
the global issue of stabilization of moduli of extra D3-branes
which are required to saturate the $C^{(4)}$ tadpole constraint.}
The basic idea is that once the
K\"ahler moduli are stabilized by
the D3 instanton contributions,
they can play the role of FI parameters
for the $U(1)_B$ gauge symmetries;
the resulting D-term potentials, which
are finite in the compact model, then
lift the runaway directions in open-string field space.

The full superpotential for $\CN=1, D=4$ supergravity coupled to
the quiver gauge theory has the form
$$W_{total}=W_{flux}+W_{tree}+W_{non-pert}+
y^a\bigl(ZV^a-\half X_{ib}X_{jc}
\epsilon^{abc}\epsilon^{ij}\bigr) $$
where $y^a, a=1,2,3$ are Lagrange multipliers. The added constraints
reflect the fact that combinations $Z,V^a,X_{ib}$
of basic quiver fields $Q,u^i,L^a$ are not independent
\eqn\gaugeinvcomb{
Z={\rm det}_{fj} Q^f u_j,\quad X_{ia}=Q u_i L_a,\quad V^a=\half L_b L_c \epsilon^{abc} ~.}
Here $W_{flux}$ is the contribution due to background fluxes
which is
responsible
for stabilization of complex structure moduli. $W_{tree}$ is the tree level superpotential in
quiver gauge theory. Finally,
$W_{non-pert}$ includes D3-instanton corrections $W_{inst}=W_{ADS}+W_{\rm stringy}$ computed
in this paper as well as $W_{gaugino},$
 the contribution from
gaugino condensation in pure SO(8) gauge theory supported on the divisor at infinity $D_6.$\foot{This is model dependent, and assumes we did not choose to put a
nontrivial vector bundle in this $SO(8)$ to absorb some of the D3 charge
tadpole.}
This divisor is a fixed set under orientifold action and therefore an O7-plane
is wrapped on it together with 4 D7-branes. Since $D_6$ does not intersect
$D_5,$ where the quiver branes are wrapped, this particular contribution does not contain operators from the quiver gauge theory, and
is just given by\foot{We ignore possible dependence on complex structure moduli.}
$$W_{gaugino}\sim \Lambda^3_{SO(8)} e^{-\alpha S_3}, $$
where $\alpha$ is a number less than one.
Here
\eqn\defsiii{S_3=\int_B \left({1\over 2\pi g_s} J^2+iC^{(4)} \right)D_6.}
Parametrizing the K\"ahler form as in previous sections: $J=r_5D_5+r_2D_2+r_3D_3$ gives
$${\rm Re}(S_3)=r_2r_3-\half r_3^2.$$
Collecting all the non-perturbative corrections we find
\eqn\mainresult{
W_{non-pert}=\Lambda^7\Biggl({1\over Z}e^{-S_1}+{1\over M_s^6}
\sum_{b>1 \& b \le -1} V_3 f(b) e^{-S_1+(b-1)S_2}\Biggr)+\Lambda^3_{SO(8)} e^{-\a S_3}.}

\noindent Let us denote $\rho_a={\rm Re}(S_a).$ The regime of validity of \mainresult\
is
\eqn\validity{\rho_3 \gg \rho_1 \gg 1,\quad \vert S_2 \vert \ll 1.}
The conditions on $S_1$ and $S_3$ are imposed to justify
neglecting higher exponentials of the type $e^{-nS_1}$ for $n>1$.
Please note that we are forced to require $\rho_3 \gg \rho_1$ since
this is the only way to have $\rho_3\gg 1,\, \rho_1 \gg 1$ and, at the same time,
ensure that sizes of the 2-cycles $C_0,f$ as well as of the $\IP^1$-fiber are non-negative:
$$r_2\ge r_3+r_5,\quad r_3\ge 2r_5,\quad r_5\ge 0.$$
Meanwhile, the condition on $S_2$ is the analytic continuation of the
condition for the alignment of the three quiver nodes discussed in \S5.2.

The fields $Z$ and $X_{ia}$ can be integrated out in the same manner as in
\IntriligatorAW\ and the effective superpotential takes the form
\eqn\effectivesuper{
W_{eff}=W_{flux}+
3\biggl(h^2\, \Lambda^7\, V_3 e^{-S_1}\biggr)^{1/3}+
 {\Lambda^7\over M_s^6}
\sum_{b>1 \& b \le -1} V_3 f(b) e^{-S_1+(b-1)S_2}+\Lambda_{SO(8)}^3 e^{-\a S_3}.}
Note that there is no susy vacuum in the regime of validity \validity\
of $W_{eff}$ since,
for example, the equations $D_{S_2}W_{eff}=0$ and $D_{V_3} W_{eff}=0$
are incompatible with each other. Indeed, in the regime \validity,\
$$\k^2 \partial_{V_3} K\sim {\rho_1\over \rho_3}\ll 1, \, ~~
\k^2 \partial_{S_2} K\sim {1 \over \sqrt{\rho_3}}\ll 1.$$
Therefore,
$D_{S_2}W_{eff}\sim \p_{S_2}W_{eff}$ and $D_{V_3} W_{eff} \sim \p_{V_3}W_{eff}$
and the equations $\p_{S_2}W_{eff}=0$ and $\p_{V_3} W_{eff}=0$ are incompatible
with each other.

To find out if there is a stable minimum one should
have a closer look at the full scalar potential.
Besides the superpotential $W_{eff}$
we need the K\"ahler potential. We would like to emphasize
that the regime of validity \validity\ implies
\eqn\valid{
r_2\sim (3/2)r_3 \gg r_5 \gg  1}
which is consistent with taking the
large radius limit. So one may use for the K\"ahler
potential the standard large radius expression:
\eqn\kahlerpot{\k^2 K=-2{\rm log}\Bigl({\rm vol}_B+
{{\rm vol}_{D_5}\over M_{pl}^2} \sqrt{V_aV_a^*}\Bigr),\quad \k^2={1\over M_{pl}^2}.}
Here ${\rm vol}_B$ is the volume of the threefold $B$ in string units,
$${\rm vol}_B={1\over 3!}\int_{B} J^3={1\over 3!}\Bigl(r_5^3-3r_3r_5^2-3r_5r_3^2
-6r_2r_5^2+6r_2r_3r_5\Bigr)$$
and ${\rm vol}_{D_5}$ is the dimensionless volume of the divisor $D_5$
$${\rm vol}_{D_5}=\half(8r_5^2-r_3^2-2r_3r_5-4r_2r_5+2r_2r_3)$$
where $r_2,r_3,r_5$ are expressed in terms of
$$\rho_a={\rm Re}(S_a),~~ a=1,2,3$$
by using the definitions of $S_a$ given in \defsi,\defsii\ and \defsiii.
In writing \kahlerpot\ we have used the kinetic term\foot{
Recall that $ds_{10}^2={\rm vol}_B^{-1}ds_4^2+ds_6^2$.} for $V_a$
$$ \int_{M^4} \sqrt{g} {{\rm vol}_{D_5}\over {\rm vol}_B} g^{\mu \nu} \p_{\mu}V_a \p_{\nu}V_a^{*}.$$

Equipped with the superpotential $W_{eff}$ and the K\"ahler potential $K$, one may use the standard formula for
the scalar potential in ${\cal N}=1,D=4$ supergravity:
\eqn\potential{V={\rm exp}(\k^2 K)
\Biggl(K^{I\Jb}W_{eff ;I}W^*_{eff ; \Jb}-3\k^2W^*_{eff}W_{eff}\Biggr)
+{1\over 2 g_X^2}\sum_{a=1}^3 (D_a)^2}
where the $U(1)$ D-terms are given by:
$$D_1=-D_2=-2 \Bigl(V^a\p_{V^a}K+\p_{\rho_1}K\Bigr),\quad D_3=0.$$
We have used the fact that both $V^a$ and $e^{S_1}$ have $U(1)$ charges $(-2,2,0)$ and
$e^{S_2}$ and $e^{S_3}$ are neutral.

It is a hard technical problem to analyze
the dynamics of the six complex scalars $V_1,V_2,V_3,S_1,S_2,S_3$ which enter in \potential.
However, we now propose a strategy
to show
that a minimum of \potential\ does exist. If $W_{flux}$ is large, one may write
$W_{eff}=W_{flux}+W_{correction}$ where
$$W_{correction}=3\biggl(h^2\, \Lambda^7\, V_3 e^{-S_1}\biggr)^{1/3}+
 {\Lambda^7\over M_s^6}
\sum_{b>1 \& b \le -1} V_3 f(b) e^{-S_1+(b-1)S_2}+\Lambda_{SO(8)}^3 e^{-\a S_3}.$$
Since $W_{flux}$ is large  it is reasonable to keep only terms where $W_{correction}$
appears no more than once. Also, for ${\rm vol}_B$ large but finite,
by adjusting $W_{flux}$ one can make
 ${1\over {\rm vol}_B}\vert W_{flux}\vert^2$ much larger than the D-terms and discard the latter. In this regime one may approximate $V_{eff}$ as follows
$$
{1\over {\rm vol}_B}\Biggl[
\Bigl(-3\k^2+K^{I\bar J}\p_I(\k^2 K) \p_{\bar J}(\k^2 K)\Bigr)
\vert W_{flux}\vert^2
+ K^{I \bar J} \bigl[ \p_{I}(\k^2 K)W_{flux}\p_{\bar J}(\bar W_{correction})+cc\bigr]\Biggr].$$

Now, in order to argue for the existence of a minimum of the theory,
it is sufficient to check two things:

\noindent
1. Show that dynamics of axions ${\rm Im} S_a$ and phases ${\rm arg} V_a$
are such that they stabilize
(for fixed values of ${\rm Re} S_a$ and $\vert V_a \vert$)
at values for which  mixed terms in $V_{\rm eff}$, which involve $W_{flux}$ and $W_{correction},$
are negative\foot{The sign of these mixed terms depends on axions and phases.}.

\noindent
2. Check that $-3\k^2+K^{I\bar J}\p_I(\k^2 K) \p_{\bar J}(\k^2 K)\le 0$
 in the regime of validity \valid\ of $W_{\rm eff}.$

If both conditions are achieved, the potential
is negative at some configuration inside the regime of validity.
Recall now that in the limit
${\rm vol}_B,{\rm vol}_{D_5} \rightarrow \infty$ (also in the regime of validity)
$V_{eff}\rightarrow 0.$
Since $V_{eff}$ is bounded below (even outside our regime of control, as
long as the full model is well-defined),
there is a minimum somewhere, though perhaps outside of the regime of
validity of $V_{eff}.$ It would be interesting to explore the conditions under which one can find
minima with exponentially small supersymmetry breaking due to the dynamics
of the quiver gauge theory.

Also, if $\vert V_3 \vert$ is large enough
$${\k^2 \vert W_{flux}\vert^2\over {\rm vol}_B}\ll  {\vert V_3\vert^2 {\rm vol}^2_{D_5}\over
{\rm vol}^2_B}, $$
the D-terms dominate over the F-term
and the potential at infinity ${\rm vol}_B, {\rm vol}_{D_5} \rightarrow \infty$
is positive (and approaching zero).
Combined with $V_{eff}<0$ at some configuration inside the regime of validity this suggests existence of a dS maximum in the regime of validity
of $V_{\rm eff}.$

\appendix{A}{Mixed Anomalies}
There is mixed anomaly $U(1)_R U(1)_a U(1)_b$
in the theory with baryonic $U(1)_B$ factors included:
\eqn\anom{
\del_\mu j^\mu_R = -2 \lambda_R r_a r_b \star (F_a\wdg F_b) .}
Here $a,b=1,2,3$ label the nodes of the quiver and the gauge field strength $F_a$ of
the $U(1)_a$ factor is a 2-form in $\IR^4.$

The compact axions $\Phi,\phi_{C_0},\phi_f$ should not transform
under R-symmetry since they couple to quiver gauge theory as:
$$I_{comp.}={\rm tr}_1 F^2  \Phi+
 {\rm tr}_2 F^2 \Biggl(\Phi+\phi_{f}+\phi_{C_0}\Biggr)
-{\rm tr}_3 F^2 \Biggl(\Phi+\phi_{f}\Biggr).
$$
Any shift of $\Phi,\phi_{C_0},\phi_f$ would result in generating
another anomaly $U(1)_R SU(N)^2$ which is otherwise absent in the
theory.

We propose a mechanism for cancellation of an $U(1)_R U(1)_a U(1)_b$ anomaly
which involves ${\hat C}_2,$ the RR 2-form in $\IR^4$ and $C'_2,$ the 2-form in $\IR^4$ obtained from
KK reduction of the RR 4-form:
$$C_4=C'_2\wdg\omega_f+\ldots .$$
The specific KK component $C'_2$ is chosen due to the property that
it couples only to $F_2$.

These 2-forms couple to the gauge fields as follows
\eqn\noncomp{I_{non-comp}={1\over 12} \sum_{a=1}^3 r_a \int_{M^4}
{\hat C}_2 \wdg F_a +\half \int_{M^4} {\hat C}_2 \wdg F_2 +
2 \int_{M^4} C'_2 \wdg F_2.}
Now we require the following non-trivial transformations under R-symmetry:
\eqn\rtransf{ \delta_R {\hat C}_2=24 \lambda_R \sum_{a=1}^3 r_a F_a,\quad
\delta_R C'_2=-6 \lambda_R \sum_{a=1}^3 r_a F_a.}
As a result of these transformations
$$\delta_R I_{non-comp}=2 \lambda_R r_a r_b F_a\wdg F_b$$
and the anomaly \anom\ is completely cancelled.
%



\break
\bigskip
\bigskip
\centerline{\bf{Acknowledgements}}
\medskip
We thank Riccardo Argurio, Matteo Bertolini, Chris Herzog, Ken Intriligator, Anton Kapustin, Igor Klebanov, Melissa Liu, Liam McAllister, Hirosi Ooguri, Eva Silverstein,
Peter Svrcek, Jay Wacker, Johannes Walcher and Martijn Wijnholt for helpful discussions.
The work of the Stanford authors was supported in part
by the NSF under grant PHY-0244728,
the DOE under contract DE-AC03-76SF00515, and a David and Lucile
Packard Foundation Fellowship for Science and Engineering.
The work of N.S. was supported in part by NSF grants
PHY-0244821 and DMS-0244464.
The work of J.M. is supported in part by funds provided
by the U.S. Department of Energy (D.O.E.) under cooperative
research agreement DE-FG0205ER41360.
S.K. is grateful to the KITP for hospitality.

\listrefs
\end